\documentclass[aps,prl,twocolumn,groupedaddress]{revtex4}

\newcommand{\YPrBCO}{Y$_{1-x}$Pr$_{x}$Ba$_{2}$Cu$_{3}$O$_{6.97}$}

\newcommand{\BSCCO}{Bi$_{2}$Sr$_{2}$CaCu$_{2}$O$_{8}$}

\newcommand{\YBCOx}{YBa$_{2}$Cu$_{3}$O$_{6.5}$}

\newcommand{\YBCO}{YBa$_{2}$Cu$_{3}$O$_{7-\delta}$}

\newcommand{\SCCO}{Sm$_{1.85}$Ce$_{0.15}$CuO$_{4-y}$}

\newcommand{\HgT}{H$_{g}(T)$}

\newcommand{\BI}{Blatter and Ivlev}

\newcommand{\Tc}{T$_{c}$}

\usepackage{graphicx}
\usepackage{epstopdf}

\begin{document}


\title{Universal Critical Dynamic Form of  the Vortex-Lattice Melting Line}


\author{B. J. Taylor}
\affiliation{Department of Physics and Institute for Pure and Applied
Physical Sciences,\\
University of California, San Diego, La Jolla, CA 92093}

\author{M. B. Maple}
\affiliation{Department of Physics and Institute for Pure and Applied
Physical Sciences,\\
University of California, San Diego, La Jolla, CA 92093}

\date{\today}

\begin{abstract}
A modified form of the vortex-lattice melting line is arrived at by incorporating the effects of critical behavior at the melting transition.  Beginning with the universal form established by \BI~ [Physical Review Letters $\bf{70}$, 2621 (1993)] which includes both thermal and quantum fluctuations, we then use the vortex relaxation time, $\tau_{r}$, of a vortex-glass with a finite transition temperature that follows from the scaling theory of Fisher, Fisher, and Huse [Physical Review B, $\bf{43}$, 130 (1991)].
This new form of the melting line is demonstrated to fit over the \textit{entire} melting line of \YPrBCO~  (x = 0 - 0.4) samples within the temperature range 0.03 $T_{c} \leq T \leq T_{c}$ \mbox{($H \leq$ 45 tesla)}  implying no crossover in dynamics from 3D to 2D.
Generically it can be seen that a change in dimensionality of the vortex fluctuations along the melting line must be accompanied by a corresponding change in the critical exponents $\nu$ and $z$. Such a change is observed for the highly anisotropic cuprate superconductor \BSCCO. 

\end{abstract}


\maketitle



Since the discovery of the high-temperature layered cuprate superconductors, the physics of the mixed state of strongly type-II superconductors has received considerable attention. A significant amount of effort has been devoted to developing a consistent theory of the melting transition of the vortex lattice.  
As there are many processes to be taken into consideration; thermal and quantum fluctuations, pinning mechanisms, correlated and un-correlated disorder, anisotropy, and  coupling of the vortex lattice to the underlying electronic structure; arriving at an expression that is relevant over the entire vortex lattice melting line has proven elusive.
In this Letter, we combine two of the most well known theories in the physics of vortex dynamics: (1) The universal form of the vortex lattice melting line of \BI~ \cite{Blatter93,Blatter94}, in which quantum fluctuations are taken into account along with the classical thermal result from the Lindemann criterion; and (2) the vortex glass critical dynamics and scaling theory of Fisher, Fisher, and Huse (FFH) \cite{Fisher91}, via the intrinsic vortex relaxation time, $\tau_{r}^{v}$, expressed through the frequency dependent form of the melting line of a vortex glass with a finite melting temperature, T$_{g}$ \cite{Fisher91,Gammel90}. The resulting expression is expected to be valid over the \textit{entire} H$_{g}(T)$ melting line, with any changes in the critical behavior of the vortex lattice along the melting transition reflected in the value of the critical exponent $\nu(z + 2 - d)$. We use the terms vortex lattice and vortex glass interchangeably due to the different terminology used in \cite{Blatter93,Blatter94} and \cite{Fisher91}; however, the expression of the melting line derived below is that of a second order vortex glass transition.

We show that this new expression does in fact describe the \textit{entire} vortex glass melting line of \mbox{\YPrBCO~ (x = 0 - 0.4) thin film samples}, \YBCOx~ single crystal samples, \mbox{(0.03 $ \leq T/T_{c} \leq$ 1)} \cite{Taylor05}, and the electron-doped cuprate \SCCO~ \cite{Scanderbeg05}, with a constant value of the critical exponent. However, the melting line of the highly anisotropic  superconductor \BSCCO, along which the vortex lattice undergoes a well known 3D to 2D transition at \mbox{H $\approx$ 1 kOe} \cite{Schilling93}, exhibits a change in the critical exponent from that of a 3D-XY like to a \mbox{2D-XY} like value at this field.

\begin{figure}
 \includegraphics[scale=0.29]{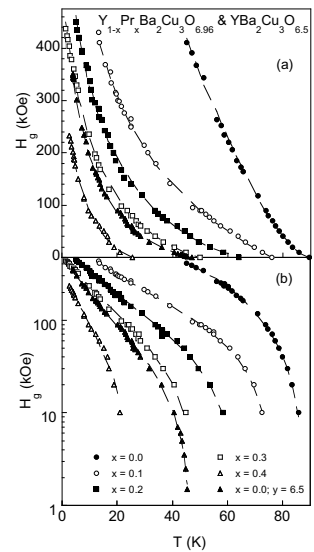}
 \caption{Vortex melting line data for \YPrBCO~ films (1kOe $< H <$ 450 kOe) and a \YBCOx~ single crystal (100 Oe $< H <$ 450 kOe) with fits of the modified melting line equation (\ref{Hm(T):New}) shown in linear (a) and semi-log (b) plots to emphasize the quality of the fits over the entire $H$ and $T$ range.\label{YBCO}}
 \end{figure}

The vortex glass melting line, \HgT, of the \YPrBCO~thin film samples \mbox{(x = 0 - 0.4)} and ultra high purity oxygen deficient \YBCOx~ single crystal, were investigated in magnetic fields up to 45 tesla. The  \HgT~line was determined from electrical resistivity, $\rho(H,T)$, measurements with H $\parallel$ c.  Low field measurements, \mbox{H $\leq$ 9 T},  were made at UCSD, whereas high field measurements were made at the National High Magnetic Field Laboratory (Tallahassee, Fl).  Further experimental details will be included in a separate publication \cite{Taylor05}. The H$_{g}(T)$ data are shown in figure (\ref{YBCO}) in linear and semi-log plots, to emphasize the quality of the fits to the data by the melting line equation derived below.

In their seminal work, \BI~ \cite{Blatter93,Blatter94} included the contribution of quantum fluctuations to the statistical mechanics of the vortex system of a type II superconductor.  The scope of the classical formalism based upon the continuum elastic theory for the vortex lattice was extended to a dynamic formalism based on the Euclidean action. Combining this new theoretical framework with the Lindemann criterion they establish a universal form of the melting line, $H_{m}(T)$ \mbox{($\equiv$ H$_{g}(T)$)}. 

It should be noted that two slightly different expressions are arrived at in \cite{Blatter93} and \cite{Blatter94}. In their initial work \cite{Blatter93}, when calculating the mean squared displacement amplitude, $\left<u^{2}\right>$, a term involving compressional modes is dropped. In the latter work \cite{Blatter94} it is not. For the first case, they obtain 
\begin{equation} \label{HmBlatter1}
H_{m} = \frac{4H_{c2}(0)\theta^{2}}{(1+\sqrt{1 + 4Q\theta})^{2}} 
\end{equation}
\\
where $\theta$ is a reduced temperature given by \mbox{$\theta = (\pi c_{L}^{2}/\sqrt{G_{i}})(1 - t)$}, $Q = [\tilde{Q}_{u}/(\pi^{2}\sqrt{G_{i}})]\Omega\tau_{r}$ is a parameter measuring the relative strength of quantum to thermal fluctuations, $\tilde{Q}_{u} = \frac{e^{2}}{\hbar}\frac{\rho_{N}}{d}$ is the dimensionless quantum resistance, $c_{L}$ is the Lindemann number, $G_{i} = [T_{c}/H_{c}^{2}(0)\epsilon\xi^{3}(0)]^{2}/2$ is the Ginzburg number, $\Omega$ is a cutoff frequency, $\tau_{r}$ is the scattering relaxation time of the quasiparticles in the vortex core given by the Drude formula $\sigma_{N} = e^{2}n\tau_{r}/m$ 
($\sigma_{N}$ is the normal state conductivity, $n$ is the free-carrier density, and $m$ the electron mass), and $d$ is the distance between the superconducting planes. $\Omega$ is given by  \mbox{$\Omega =$ min[$\Omega_{\rho}, \Omega_{i}$]}, where $\Omega_{\rho} \approx \sqrt{\eta_{\ell}/\mu_{\ell}\tau_{r}}$ is the kinetic cutoff frequency associated with the electromagnetic contribution to the vortex mass, and $\Omega_{i} \approx \frac{2}{\hbar}\Delta$ is the intrinsic cutoff given by the gap energy due to the creation of quasiparticles by vortex motion.  For the latter case, they find
\begin{equation} \label{HmBlatter2}
H_{m} = \frac{4H_{c2}(0)\theta^{2}}{(1+\sqrt{1 + 4S\theta/t})^{2}} 
\end{equation}
where now $\theta = c_{L}^{2}\sqrt{\frac{\beta_{th}}{G_{i}}}(T_{c}/T - 1)$, $S = q + c_{L}^{2}\sqrt{\frac{\beta_{th}}{G_{i}}}$, and $q = \frac{2\sqrt{\beta_{th}}}{\pi^{3}}\frac{Q_{u}}{\sqrt{G_{i}}}\Omega\tau_{r}$.
Either expression can be approximated by the power-law form H$_{m} \sim (1- t)^{\alpha}$ over temperatures ranging from \Tc~ down to T $\approx$ 0.6 \Tc. By estimating values for $\tilde{Q}_{u}$ and $\sqrt{G_{i}}$ and leaving $c_{L}$ and $\Omega\tau_{r}$ as fitting parameters in \cite{Blatter93}, 
they find $\alpha \approx$ 1.45, in close agreement with experimental values. 
As pointed out by \BI, the value of the approximate exponent $\alpha$ depends on the quantum parameter $Q$ and the reduced temperature $\theta$. This then explains the experimentally observed increase of  $\alpha$ as the temperature drops below \mbox{T $\approx$ 0.6 \Tc} \cite{Almasan92}.  

Instead of putting in constant values or approximations for the various factors of the quantum parameter $Q$, we leave them in their exact temperature and field dependent forms. Additionally, we allow for two intrinsic relaxation times; one given by the quantized level splittings of the quasiparticle states within the vortex core \cite{Caroli65,Karrai92}, $\tau_{r}^{core}$, and the second given by the motion of the vortex $\tau_{r}^{v}$, with $\tau_{r}^{v} \gg \tau_{r}^{core}$ as the melting transition is approached \cite{Sagdahl90}. 

With $\sigma_{N} \propto \tau_{r}^{core}$, then we identify $\tau_{r}^{v}$ as the relevant relaxation time to be associated with the cutoff frequency $\Omega_{\rho}  \approx \sqrt{\eta_{\ell}/\mu_{\ell}\tau_{r}^{v}}$ and the expression for the quantum parameter $Q = \frac{\tilde{Q}_{u}}{\pi\sqrt{G_{i}}}\Omega\tau_{r}^{v}$. The divergent behavior of $\tau_{r}^{v}$ then leads to $\Omega_{\rho} < \Omega_{i}$. (For \YBCO~ \BI~ find $\Omega_{\rho} \sim 10~\Omega_{i}$ using the assumption $\tau_{r}^{v} = \tau_{r}^{core}$ \cite{Blatter94}). The vortex relaxation time $\tau_{r}^{v}$ in the critical region of the melting transition is expressed through the frequency dependent form of the melting line from the critical scaling theory of \mbox{FFH}. This choice is appropriate as we are seeking an expression relevant \textit{in the region of melting} of the lattice. This temperature and field dependent  form of $Q = Q[H, t]$ (or $q = q[H, T]$), inserted back into equation (\ref{HmBlatter1}) (or (\ref{HmBlatter2})), results in an equation that ties together the dynamics of thermal and quantum fluctuations with the critical behavior of the vortex-glass at the melting transition. It readily follows then that a deviation from this melting, i.e, a 3D - 2D transition, must be accompanied by a corresponding change in the critical exponent $s \equiv  \nu(z + 2 - d)$.
 \begin{table}
 \caption{Values of the Lindemann number $c_{L}$, critical exponent $s \equiv \nu(z +2 - d)$, and the quantum parameters $Q_{0}$ or $q_{0}$ using equation (\ref{Hm(T):New})$^{\dagger}$ or equation (\ref{HmBlatter2}) with (\ref{q(Tg)})$^{\ddagger}$ for the data in figure \ref{YBCO}.\label{parameters}}
 \begin{tabular}{|c|ccc|ccc|} \hline \hline
  ~~~x~~~ & ~~~c$_{L}^{\dagger}$~~~& ~~~$s^{\dagger}$~~~& ~~~$Q_{0}^{\dagger}$~~~&~~~c$_{L}^{\ddagger}$~~~& ~~~$s^{\ddagger}$~~~& ~~~$q_{0}^{\ddagger}$~~~\\
\hline 
 0~  & 0.31 & 3.33 & 0.34 & 0.34 & 4.6 & 0.1\\
 0.1 & 0.28 & 1.90 & 11.5 & 0.31 & 1.2 & 6.7\\
 0.2 & 0.29 & 2.07 & 12.8 & 0.29 & 1.8 & 2.0\\  
 0.3 & 0.30 & 2.10 & 9.8   & 0.30 & 2.0 & 2.3\\
 0.4 & 0.27 & 2.22 & 9.6   & 0.31 & 1.8 & 1.2\\ \hline
 y=6.5& & & & & & \\ \hline
 0 & 0.28 & 2.21 & 16.9 & 0.21 & 1.2 & 1.2\\ \hline
 \end{tabular}
 \end{table}

Starting with the expression for the kinetic cutoff frequency, using the Bardeen-Stephen expression for the viscous drag coefficient \cite{Bardeen65},
\begin{equation}\label{Bardeen}
 \eta_{\ell} \approx \frac{\Phi_{0}^{2}\sigma_{N}}{2\pi \xi^{2}},
\end{equation}
using the electromagnetic contribution of the vortex mass \cite{Suhl65}, 
\begin{equation} 
\mu_{\ell}^{em} = \frac{1}{4\mu_{0}}\frac{\xi^{2}H_{c}^{2}}{c^{2}}\left(\frac{\lambda}{\lambda_{d}}\right)^{2}
\end{equation}
and including the temperature dependence of \mbox{$\xi = \xi_{0}/(1 - t)^{\frac{1}{2}}$}, we then have
\begin{equation} \label{Omega(t)}
\Omega \approx \sqrt{\frac{\eta_{\ell}}{\mu_{\ell}^{em}\tau_{r}^{v}}} = \frac{4c\lambda_{d}}{\xi_{0}}\sqrt{\frac{\pi\mu_{0}\sigma_{N}}{\tau_{r}^{v}}} \left(1-t\right)^{\frac{1}{2}},
\end{equation}
where $\lambda_{d}$ is a shielding length, which is a few  times that of ($k_{F}$)$^{-1}$.

Within the FFH critical dynamic theory of a thermodynamic vortex-glass transition \cite{Koch89,Fisher91}, physical properties are expressed through a diverging correlation length $\xi_{VG} \sim |T - T_{g}|^{-\nu}$ and correlation time $\tau_{VG} \sim \xi_{VG}^{z}$. From the scaling form of the frequency dependent conductivity $\sigma(\omega) \approx \xi_{VG}^{z + 2 - d}\textit{S}_{\pm}(\omega\xi_{VG}^{z})$, the vortex relaxation rate for a glass model with a finite transition temperature expressed as a frequency is found to be \cite{Gammel90},
\begin{equation} \label{omegam}
\omega_{m} = \left(\frac{\omega_{0}}{T_{g}^{s}}\right)\left(T_{c}(\omega) - T_{g}\right)^{s}.
\end{equation}
where $s \equiv \nu(z + 2 - d)$ and $d$ is the dimensionality of the vortex glass system.
Returning to a relaxation time, we have,
\begin{equation} \label{taur(t)}
\tau_{r}^{v} \equiv \tau_{VG} =  \tau_{0}\left(\frac{T_{g}}{T_{c}}\right)^{s} \left(1- \frac{T_{g}}{T_{c}}\right)^{-s},
\end{equation}

Combining (\ref{taur(t)}) with (\ref{Omega(t)}), with all expressions evaluated at the melting temperature $T_{g} \equiv T_{m}$, we have 
\begin{equation} \label{Omega(Tg)}
\Omega = \frac{4c\lambda_{d}}{\xi_{0}}\sqrt{\frac{\pi\mu_{0}\sigma_{N}}{\tau_{0}}}\left(\frac{T_{g}}{T_{c}}\right)^{-s/2} \left(1- \frac{T_{g}}{T_{c}}\right)^{(1+s)/2}
\end{equation}

Next we use the expression for the dimensionless quantum of resistance given in \cite{Blatter94a},
\begin{equation}\label{Qu}
\tilde{Q}_{u} = \frac{e^{2}}{\hbar}\frac{\rho_{N}}{\epsilon\xi},
\end{equation}
and the field dependent expression of the Ginzburg number \cite{Bennemann03}, 
\begin{equation}\label{G(H)}
G_{i}(H_{g}) \approx \left(G_{i}\right)^{\frac{1}{3}} \left(\frac{H_{g}}{H_{c2}(0)}\right)^{\frac{2}{3}}.
\end{equation}
Combining (\ref{taur(t)}), (\ref{Omega(Tg)}), (\ref{Qu}), and (\ref{G(H)}),  we arrive at the final expression for the quantum parameters $Q$ and $q$,

\begin{equation} \label{Q(Tg)}
Q(H_{g}, T_{g}) = \frac{\tilde{Q}_{0}\Omega_{0}\tau_{0}}{\pi^{2}\sqrt{G_{i}(H_{g})}}t^{\frac{s}{2}}(1-t)^{1-\frac{s}{2}},
\end{equation}
\begin{equation} \label{q(Tg)}
q(H_{g}, T_{g}) =  \frac{2\sqrt{\beta_{th}}}{\pi^{3}} \frac{\tilde{Q}_{0}\Omega_{0}\tau_{0}}{\sqrt{G_{i}(H_{g})}}t^{\frac{s}{2}}(1-t)^{1-\frac{s}{2}},
\end{equation}
with $\Omega_{0} \equiv  \frac{4c\lambda_{d}}{\xi_{0}}\sqrt{\frac{\pi\mu_{0}\sigma_{N}}{\tau_{0}}}$ and $\tilde{Q}_{0} \equiv \frac{e^{2}}{\hbar}\frac{\rho_{N}}{\epsilon\xi_{0}}$. These can then be inserted back into the melting line equations (\ref{HmBlatter1}) and (\ref{HmBlatter2}). Combining (\ref{HmBlatter1}) with (\ref{Q(Tg)}), we have
\begin{equation} \label{Hm(T):New}
H_{g}(Q) = \frac{4H_{c2}(0)\frac{(\pi c_{L}^{2})^{2}}{G_{i}(H_{g})}(1-t)^{2}}{\left(1 + \sqrt{1 + 4(\tilde{Q}_{0}\Omega_{0}\tau_{0})\frac{c_{L}^{2}}{\pi G_{i}(H_{g})} t^{\frac{s}{2}} (1-t)^{2-\frac{s}{2}}}\right)^{2}},
\end{equation}
\begin{figure}
 \includegraphics[scale=0.28]{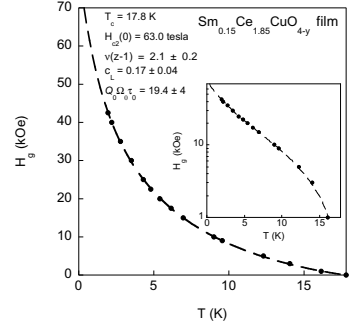}
 \caption{Fit of equation (\ref{Hm(T):New})  to the melting line, $H_{g}(T)$,
  of a thin film of the electron-doped cuprate superconductor \SCCO~ \cite{Scanderbeg05}. 
  Inset is the same data on a semi-log plot.\label{SCCO}}
 \end{figure}
\begin{figure}
 \includegraphics[scale=0.28]{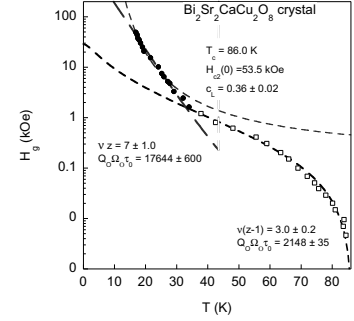}
 \caption{Fit of equation (\ref{Hm(T):New}) indicating a 3D - 2D transition at H$_{2D} \approx$ 1 kOe. A corresponding change of the exponent $\nu(z + 2 - d)$ from a 3D-XY like value to that of a 2D glass is observed. The upper portion of the curve is fit to the 2D melting line expression of Schilling et al. \cite{Schilling93}  for comparison.  \label{BSCCO}}
 \end{figure}
%
It should be noted that the expression used for the field dependent Ginzburg number in (\ref{G(H)}) is arrived at from a melting line that follows the power-law form \mbox{$H_{m} \sim \frac{H_{c2}(0)}{\sqrt{G_{i}}}\left(1-t\right)^{\frac{3}{2}}$}, and so is strictly valid only for \mbox{$T \gtrsim$ 0.6 \Tc}. In general, for a portion of the melting line that can be approximated by  \mbox{$H_{m} \sim \frac{H_{c2}(0)}{\sqrt{G_{i}}}\left(1-t\right)^{\alpha}$}
\begin{equation} \label{G(Hn)}
G_{i}(H_{g}) \approx \left(G_{i}\right)^{\frac{1}{2\alpha}} \left(\frac{H_{g}}{H_{c2}(0)}\right)^{\frac{1}{\alpha}}. 
\end{equation}
Also, it should be recognized that at finite frequencies dispersive effects lead to \cite{Kopnin91,Blatter94}
\begin{equation}\label{etaw}
\eta(\omega) \approx \Phi_{0}\rho_{s}\frac{\omega_{0}\tau_{r}^{core}\left(1 - i\omega\tau_{r}^{core}\right)}{\left(1 - i\omega\tau_{r}^{core}\right)^{2} + \left(\omega_{0}\tau_{r}^{core}\right)^{2}}
\end{equation}
with $\rho_{s} = 2e\left|\psi\right|^{2} $ the superfluid density. Note that for large frequencies ($\omega\tau_{r} \gg 1)$ and with $\rho_{s}\sim (1 - t)$, this leads to the same temperature dependence of $\Omega$ found in (\ref{Omega(t)}). 
Alternately \cite{Blatter94,Schlesinger90}, starting from (\ref{etaw}) with the condition $\omega\tau_{r} \gg 1$, $\omega_{0}\tau_{r} \ll 1$, and using $\rho_{s} = -en$, $\omega_{o} = \omega_{c} = eB/m$, with $B \approx H_{c2}$
\begin{equation}
\eta_{\ell}(\omega) \approx \frac{\Phi_{0}^{2}}{2\pi \xi^{2}}\frac{e^{2}n\tau_{r}^{core}}{m}
\end{equation}
which is identical to the Bardeen-Stephen result (\ref{Bardeen}).
   
The melting line data, H$_{g}(T)$, obtained in this study of  \YPrBCO~ films and a \YBCOx~ single crystal were fitted by equation (\ref{Hm(T):New}) \mbox{(figure (\ref{YBCO}- a, b))} and that obtained by substitution of (\ref{q(Tg)}) into (\ref{HmBlatter2}) (not shown). With the experimental value $Q_{0} = $ 0.34 $\pm$ 0.15 obtained from equation (\ref{Hm(T):New}), we solve for the value of $\Omega_{0}\tau_{r 0}^{v}$ for the \YBCO~ film. Using $\xi_{0} = 12 \AA$, $\epsilon = 1/8$, $\rho_{N} \approx 2\times10^{-5} \Omega cm$, $\Delta \approx ~$14 meV \cite{Hirata91},  and with $\lambda_{d}$  a few times (1 - 3 ) $k_{F}^{-1}$;  \mbox{$\Omega_{0}\tau_{r0}^{v} \approx$ 0.65}.  If instead we use the second modified melting line equation (\ref{HmBlatter2}) with $\beta_{th} =$ 2.5, then we get an equally good fit indistinguishable from (\ref{Hm(T):New}) with similar values of $\nu(z -1)$ and c$_{L}^{2}$, with $Q_{0} \approx$ 0.1. In general, it is seen (see table \ref{parameters}) that the value of $q_{0}$ is reduced from $Q_{0}$ by a factor of \mbox{$\sim$ 1/2 - 1/10}. The values from both fits are listed in table \ref{parameters}. The result, $q < Q$, can be understood as a relative reduction of the quantum contribution to displacements by the inclusion of the thermal contribution of the compressional modes to the mean squared displacement amplitude of the vortices, $\left<u^{2}\right>$, used in the melting line equation (\ref{HmBlatter2}), which are not in (\ref{HmBlatter1}).

Additionally, equation (\ref{Hm(T):New}) was used to analyze melting line data from a \SCCO~ film \cite{Scanderbeg05} \mbox{(figure \ref{SCCO})}, and a \BSCCO~ single crystal \cite{Schilling93} \mbox{(figure \ref{BSCCO})}.
The data for all of the above samples are described well by equation (\ref{Hm(T):New}) over the entire range of field/temperatures examined, with the exception of the \BSCCO~ data, which can be fit by equation (\ref{Hm(T):New}) in two segments, with a corresponding change in the critical exponent $s$. This latter case is interpreted readily as evidence for a 3D-2D vortex glass transition at H$_{2D} \approx$ 1 kOe, which is well established \cite{Schilling93} in this highly anisotropic compound. Since the data in the other cases considered here can be fit over the entire (available) range of the H$_{g}(T)$ line with one set of critical exponents corresponding to 3D, this implies that the vortex glass remains in a 3D state along the entire line.

While it is $\textit{a priori}$ unclear as to which equation for the melting line is more appropriate for a given system, accurate measurements of the many physical properties (H$_{c2}$, G$_{i}$, $\xi_{0}$, etc.) that enter these equations would, in principle, make it possible to distinguish between the two.  Either way, we have arrived at a form of the melting line that unifies the quantum/thermal nature of vortex fluctuations with the critical dynamic behavior of vortices in the region of the melting line transition, providing a more complete picture of the physics involved.
\begin{acknowledgments}
This research was sponsored by the U.S. Department of Energy (DOE) under Research Grant \#
DE-FG02-04ER46105. A portion of this work was performed at the National High Magnetic Field Laboratory, which is supported by NSF Cooperative Agreement No. DMR-0084173, by the State of Florida, and by the DOE.
\end{acknowledgments}

\end{document}